\documentclass[aps,prl,preprint,showpacs,superscriptaddress]{revtex4-1}
\usepackage{graphicx,color}
\usepackage{amssymb}
\usepackage{xcolor}

\begin{document}

\title{Laser-driven plasma pinching in $e^{-}e^{+}$ cascade}

\author{E.S.~Efimenko}
\author{A.V.~Bashinov}
\affiliation{Institute of Applied Physics, Russian Academy of Sciences, 603950 Nizhny Novgorod, Russia}
\author{A.A.~Gonoskov}
\affiliation{Institute of Applied Physics, Russian Academy of Sciences, 603950 Nizhny Novgorod, Russia}
\affiliation{Lobachevsky State University of Nizhni Novgorod, 603950 Nizhny Novgorod, Russia}
\author{S.I.~Bastrakov}
\affiliation{Helmholtz-Zentrum Dresden-Rossendorf, 01328 Dresden, Germany}
\affiliation{Lobachevsky State University of Nizhni Novgorod, 603950 Nizhny Novgorod, Russia}
\author{A.A.~Muraviev}
\affiliation{Institute of Applied Physics, Russian Academy of Sciences, 603950 Nizhny Novgorod, Russia}
\author{I.B.~Meyerov}
\affiliation{Lobachevsky State University of Nizhni Novgorod, 603950 Nizhny Novgorod, Russia}
\author{A.V.~Kim}
\author{A.M.~Sergeev}
\affiliation{Institute of Applied Physics, Russian Academy of Sciences, 603950 Nizhny Novgorod, Russia}

\date{\today}

\begin{abstract}
The cascaded production and dynamics of electron-positron plasma in ultimately focused laser fields of extreme intensity are studied by
3D particle-in-cell simulations with the account for the relevant processes of quantum electrodynamics (QED). We show that, if the laser
facility provides a total power above 20 PW, it is possible to trigger not only a QED cascade but also pinching in the produced
electron-positron plasma. The plasma self-compression in this case leads to an abrupt rise of the peak density and magnetic (electric) field up to
at least $10^{28}$ cm$^{-3}$ and 1/20 (1/40) of the Schwinger field, respectively. Determining the actual limits and physics of this process might require
quantum treatment beyond the used standard semiclassical approach. The proposed setup can thus provide extreme conditions for probing and exploring fundamental 
physics of the matter and vacuum.

\end{abstract}

\maketitle

In recent years there has been an increasing interest in the problem of electron-positron-pair plasma production by laser fields of extreme intensities. This 
problem could shed light on the long-standing questions related to extreme astrophysical events, such as jets and gamma-ray bursts\cite{Meszaros2002}. The 
problem itself represents a great challenge in contemporary physics in the context of producing very dense pair plasmas in laboratory\cite{Sarri2015}. A 
pioneering step was made by Bell and Kirk who showed that 
prolific pair generation can occur in a standing 
circularly polarized wave with an intensity of about $10^{24} \mbox{W/cm}^2$ \cite{Bell2008}. The subsequent investigations of laser fields of various 
configurations confirmed the possibility of efficient generation of electron-positron pairs 
and optimized field configurations were proposed
\cite{Bulanov2010_2,Gelfer2015,Gonoskov_ART}. The next important step was done by means of 2D and 3D simulations, mostly based on particle-in-cell (PIC) codes 
extended with probabilistic routines for the processes of quantum electrodynamics (QED). The simulations showed that dense pair plasma can be 
actually produced by tightly focused laser beams \cite{Nerush2011,Grismayer_POP2016,Jirka2016}.  
Furthermore, effective $\gamma$-ray sources in the GeV energy range were also proposed \cite{Gonoskov_PRX}. Estimates show that the forthcoming 10~PW-class 
laser 
facilities, such as the Extreme Light Infrastructure (ELI) Facility\cite{ELI}, the Vulcan 10 PW upgrade \cite{VULCAN} or Apollon 10PW\cite{APOLLON} are able to 
ignite the electron-positron cascade in vacuum. It was recently demonstrated that extreme states of pair plasmas in terms of densities, currents, electric and magnetic fields can be produced in an optimally configured multi-beam laser 
setup\cite{Efimenko2018}. It should be also mentioned that along with this research a lot of fundamental issues of extremely intense laser field interactions with matter were raised and resolved, such as the processes of strong-field quantum electrodynamics (QED) 
\cite{Nikishov1967,Dipiazza2012}, the signatures of radiation reaction forces and radiation trapping effects in single-particle motion 
\cite{Gonoskov_ART,Ji2014,Fedotov_PRA2014,Cole2018,Poder2018}, and the nonlinear regimes of laser plasma interactions 
\cite{Vranic_PPCF2017,Efimenko2018,Luo2018}. However, many of the fundamental features, especially related to nonlinear interactions of 
electron-positron plasmas with laser fields of extreme intensities, are still to be studied. This is especially true in the light of the 
next-generation projects, such as the XCELS\cite{XCELS}, the Gekko-EXA\cite{Kawanaka2016} and 
the SEL\cite{SEL}, that are aimed at constructing multi-10~PW laser systems with the power of up to 100~PW. 

In this Letter we consider the cascaded production and the consequent self-action of electron-positron plasma spatially confined in a 
$\lambda^3$ volume initiated by the optimally focused 
laser radiation with a total power of multi-10 PW level. We reveal a fundamentally new feature of the field-driven electron-positron 
plasma, when an axial column of plasma current and self-generated azimuthal magnetic field give rise to plasma pinching on a time interval much less than the 
laser cycle without any numerically observable limits within the used QED-PIC treatment.

Although the formation of current structures is a fundamental property of plasma responding to external fields, 
for the problem of interest intrinsic current interactions play an exceptionally important role because of their high values. It is shown that the pinching 
effect yields extremely dense pair plasma, and has a strong impact on microscopic characteristics such as particle orbits and $\gamma$-ray emission as well as macroscopic plasma 
parameters. The rapid growth of the plasma-produced azimuthal magnetic field also causes an inductive axial electric 
field of 
extreme strength. This may provide a novel pathway for approaching the Schwinger limit despite the accepted view concerning field attainability in 
vacuum\cite{Bulanov2010,Fedotov2010}.

\begin{figure}[t!]
\centering
\includegraphics[width=8.5cm]{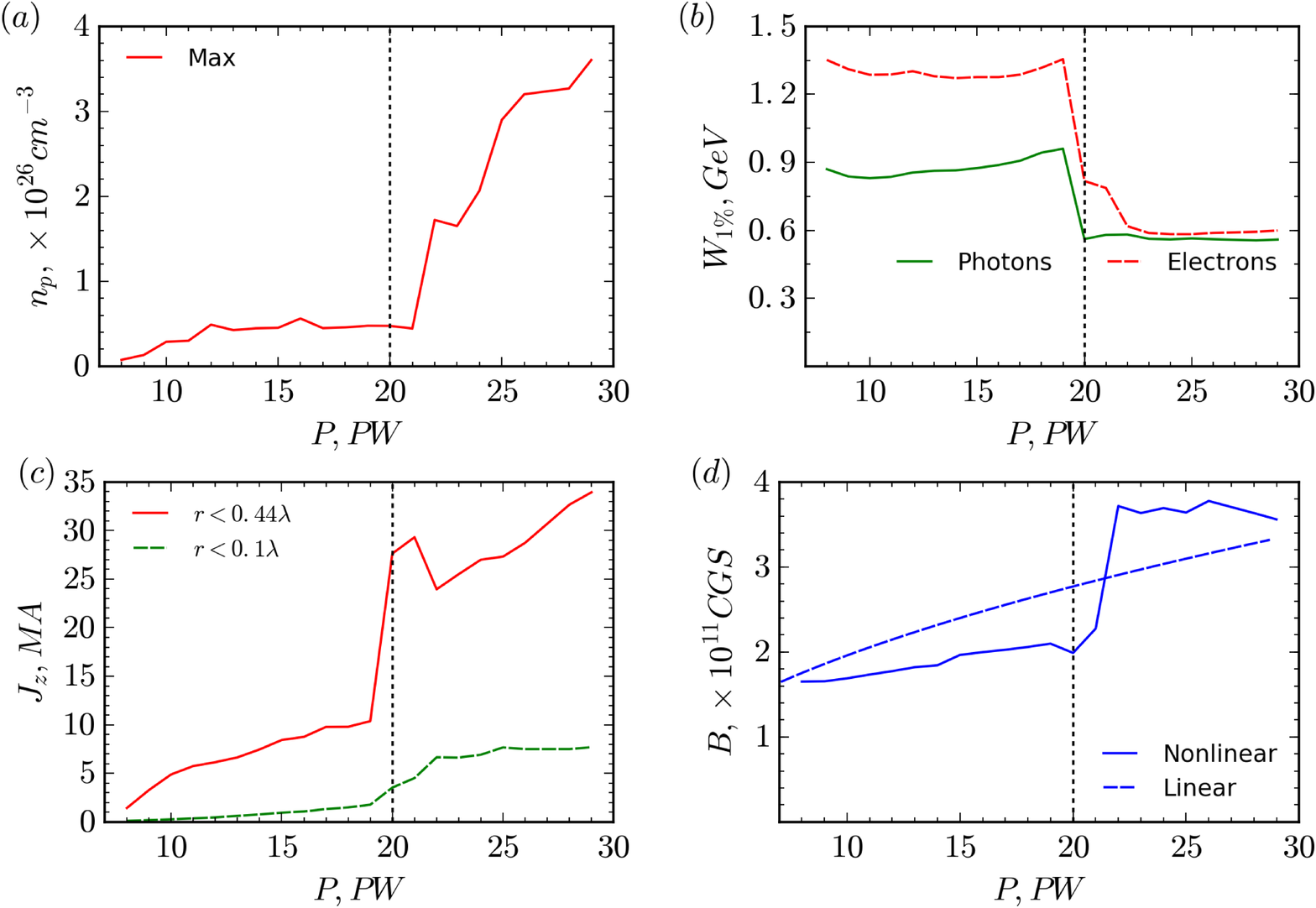}
\caption{(Color online) Plasma-field parameters at nonlinear stage of interaction {\em{vs}} laser power: (a) 
maximum pair density; (b) maximum photons (solid line) and electrons (dashed line) energy; (c) currents through plane $z$=0 in a cylinders $\rho = 0.44\lambda$ (solid line) and $\rho = 0.1\lambda$ (dashed line); (d) maximum magnetic field at 
linear (dashed line) and nonlinear (solid line) stage of interaction.
\label{fig1}}
\end{figure}

{\em{Pair plasma generation.---}}To get an insight into the physics and reduce the laser power threshold for the effect of interest we consider the ideal case 
of the optimal multi-beam configuration of the laser setup in 
the form of an incoming e-dipole wave, which minimizes the focal volume and maximizes the field 
strength\cite{Gonoskov_Dipole}. The electric field in focus is directed along the $z$-axis. We study the spatiotemporal evolution of laser-produced pair plasma 
in vacuum by performing 3D simulations with the QED-PIC 
code PICADOR\cite{Surmin2016}. The total power $P$ is varied in the range from 15 to 30 PW. The cascade is seeded by a spherical plasma target with 
electron density $10^{20}-10^{21}$~cm$^{-3}$ and a diameter of $3~\mu$m. 
The time envelope of the laser radiation is chosen to be close to rectangular with the envelope $E\{B\}(t) = E_0\{B_0\} 
\cdot f(t), f(t) = 0.25(1-\tanh(-\alpha t))(1-\tanh(\alpha (t-\tau)))$,
which has rapid fronts and a constant amplitude $E_0(B_0)$ of electric(magnetic) fields, respectively, here $\alpha=2.276/T$, the pulse 
duration $\tau = 5T$, $T$ is the laser period. The laser wavelength $\lambda = 0.9~\mu m$ is taken according to the XCELS project. A cubic grid having
size $ 4~\mu\mbox{m} \times 4~\mu\mbox{m} \times 4~\mu\mbox{m}$ and number of cells $512 \times 512 \times 512$ is used in simulations.  The time step is chosen 
to be 0.015 fs.  

This resolution is sufficient for the process of triggering the pinching effect, while the pinch itself has a singular nature and is thus analyzed using up to 
32 times higher resolution for both time and coordinate.
For modeling QED cascades an Adaptive Event Generator module\cite{Gonoskov_PRE} is used with separate particle resampling for different particle types. This 
module uses automatic time step splitting, thus photon emission and decay time are efficiently resolved during the simulations.

At a laser power exceeding the vacuum breakdown threshold, which is about 7.2 PW, pair production starts to grow exponentially in 
time \cite{Gonoskov_PRX,Efimenko2018}. The regime of interaction for $P < 20$~PW was studied in Ref.\cite{Efimenko2018}. The specific feature of this regime is 
that the electromagnetic field structure remains almost unchanged, while the amplitude is reduced.

However, for  $P>20$ PW there is a striking feature of the laser-plasma interaction that manifests itself through the rapid  and quite unexpected change of 
maximum magnetic field and plasma properties, such as maximum pair densities, maximum particle energies and maximum current 
through the central plane $z=0$ during the interaction, and indicates qualitative modification of the interaction regime.
Taking into account that in the focal region ($\rho\equiv\sqrt{x^2+y^2}<0.44\lambda$)  the electrons and positrons accelerated by the electric field form counter-streaming 
axial flows, large enhancement of plasma density (see Fig. \ref{fig1}(a)) and axial current especially within a cylinder with a small radius $\rho=0.1\lambda$ (see Fig. \ref{fig1}(c)) 
demonstrate not only an increase in the number of pairs, but a significant decrease in the radial size of the plasma structures as well. Moreover, the produced currents are 
comparable with the Alfven current $17\gamma\beta_z$ [kA]$\approx$ 20 MA, where $\gamma\approx1200$(see Fig. \ref{fig1}(b)) is the Lorentz factor of particles and 
$\beta_z \simeq 1$ is the 
axial particle velocity  normalized to the light velocity $c$.  Such a strong "wire-like" current in a narrow vicinity of the $z$-axis  can generate a magnetic 
field $2J_z/ (\rho c) \approx 10^{11}$ G close to the vacuum field $B_0\approx 10\sqrt{P}/(\lambda\sqrt{c}) \approx 10^{11}$ G and change particle 
dynamics, which is signalled by up to 1.4 excess of the magnetic field over $B_0$ (see Fig. \ref{fig1}(d)) and the drop of the  maximum particle energy (see Fig. \ref{fig1}(b)) at $P>20$ 
PW. Maximum energy $W_{1\%}$ is defined as the minimum energy of 1\% of the most energetic particles. 
All these observations have led us to the conclusion that a qualitative change of the interaction occurs as a result of the internal  plasma mechanism of magnetic field 
generation which we attribute to the formation of strong axial current.  The most appropriate process demonstrating the observed properties is 
plasma pinching, where the corresponding current-carrying plasma is significantly compressed by the self-generated magnetic field as in the 
conventional case of static external fields\cite{Benford1971}. Unlike the conventional pinch-effect, in our case electron-positron plasma  interacts with the 
oscillating laser field.

\begin{figure}[t!]
\centering
\includegraphics[width=8.5cm]{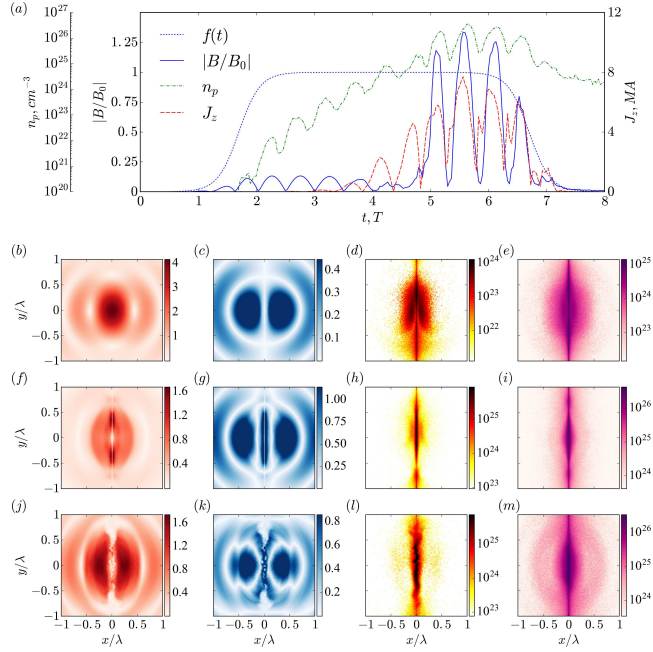}
\caption{(Color online) Plasma-field dynamics for 27 PW. (a) Temporal evolution, $f(t)$ - laser pulse envelope, $B/B_0$ - magnetic field in the plane $z = 0$ at $\rho = 1/60 \lambda$, $n_p$ - maximum pair 
density, $J_z$ - current through the cylinder with radius 0.1 $\lambda$.  Electric (b,f,j) and magnetic 
(c,g,k) fields, positron (d,h,l) and photon (e,i,m) distributions at the stages of (b-e) linear QED cascade, (f-i) 
plasma column pinching and (j-m) pinch breakdown due to bending instability.   
\label{fig2}}
\end{figure}

{\em{Pinching of plasma column.---}}  To show qualitatively different dynamics of pair plasmas let us look in detail into the spatiotemporal evolution of 
electron-positron pair plasma with a focus on highly localized plasma distribution and corresponding field structures in the vicinity of the $z$-axis. In 
Fig. \ref{fig2}(a) we present the temporal evolution for $P = 27$~PW. The spatial 
distributions of fields, positron and gamma photon densities are shown in Fig. \ref{fig2}(b-m). During the first two periods, the plasma target is compressed 
and the exponential growth of electron-positron pair plasma density is established as shown in Fig. 
\ref{fig2}(b-e). This linear regime was studied in ample detail in Ref.\cite{Gonoskov_PRX}. At this stage the magnetic field in the vicinity of the axis is low, 
which is 
explained by the $e$-dipole wave magnetic field structure with minimum in the center, see Fig. \ref{fig2}(c). Charged particles oscillate in the ART 
regime\cite{Gonoskov_ART} and emit $\gamma-$photons predominantly along the $z$-axis. The dynamics changes drastically when the current $J_z$ in the cylinder 
$\rho = 0.1\lambda$ in the 
center approaches 4-5 MA at $t=5T$, see Fig. \ref{fig2}(a). Within half of the wave period pair 
density jumps an order of magnitude and exactly at the same time there appears a huge magnetic field exceeding the unperturbed magnetic field amplitude. The 
spatial distribution clearly shows the formation of an electron-positron pair plasma column with magnetic field at its boundary, see Fig. \ref{fig2}(f-i). This 
plasma column is compressed by the self-generated magnetic field to a few cells, which leads to rapid rise in plasma density, and decay through bending 
instability, see Fig. \ref{fig2}(j-m). 
To determine whether the observed instability has a physical origin or is caused by approaching the resolution limits we perform a special series of 
simulations with up to 32 times higher spatial and 
temporal resolution , which for the finest resolution requires significant computational resources. The simulations show that the pinch dynamics, although being 
qualitatively similar, quantitatively depends on the simulation resolution. 
 
\begin{figure}[t!]
\centering
\includegraphics[width=8.5cm]{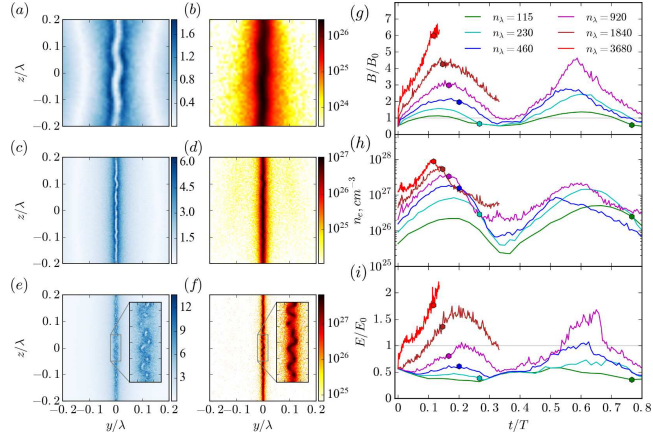}
\caption{(Color online) (a-f) Magnetic field (a,c,e) and pair density distribution (b,d,f) at the moment of maximum compression and start of bending 
instability for different resolution: (a,b) $n_\lambda = 115$, (c,d) $n_\lambda = 460$, (e,f) $n_\lambda = 1840$. (g-i) Pinch temporal evolution 
for different $n_\lambda$: (g) magnetic field, (h) pair density, (i) electric field.  Colored dots show the starting moment of bending instability.
\label{fig3}}
\end{figure}

\begin{table}[t]

 \begin{tabular}{ c | c | c | c | c | c | c | c}
 \hline
$n_\lambda$ & $n_p, \mbox{cm}^{-3}$ & $B/B_0$ & $P_B, \mbox{PW}$ & $E/E_0$ & $P_E, \mbox{PW}$ & $E_S/B$ & $E_S/E$\\ \hline
115 & $5.1 \times 10^{26}$ & 1.4 & 51 & 0.6 & 9 & 100 & 154\\
230 & $1.5 \times 10^{27}$ & 2.4 & 161 & 0.7 & 14 & 56 & 122\\
460 & $1.9 \times 10^{27}$ & 2.8 & 207 & 1.1 & 31 & 49 & 83\\
920 & $3.6 \times 10^{27}$ & 3.3 & 291 & 1.1 & 31 & 42 & 84\\
1840 & $6.5 \times 10^{27}$ & 4.6 & 583 & 1.8 & 83 & 29 & 51\\
3680 & $9.7 \times 10^{27}$ & 6.7 & 1212 & 2.2 & 132 & 20 & 40\\
 \hline
 
 \end{tabular}
 \caption{Pinch parameters for different simulation resolutions. $n_\lambda$ -- number of steps per wavelength, $n_p$ -- maximum pair density, 
$B/B_0(E/E_0)$ -- maximum magnetic(electric) field normalized to vacuum magnetic(electric) field, $P_B(P_E)$ -- power of e-dipole wave with the same value of magnetic(electric) field, $E_s/B(E_s/E)$ -- ratio of Schwinger field to maximum magnetic(electric) field.}
 \label{table_resolution}
\end{table}

In Fig. \ref{fig3}(a-f) we present spatial density distributions of the central part of the plasma column and the corresponding magnetic field. The first row 
corresponds to the initial resolution ($n_\lambda$ = 115), the second and the third rows show the same area with 4 and 16 times higher resolution($n_\lambda$ = 460 and $n_\lambda$ = 1840), respectively. Dynamics of the electric and magnetic fields and maximum pair density 
for different simulation resolutions at the stage of plasma self-compression is shown in Fig. \ref{fig3}(g-i). It can be seen that the increase of resolution leads to the  monotonic 
increase of pair-plasma density and fields, resulting in up to 7 times increase of magnetic field 
as compared to vacuum fields and pair density up to $10^{28} \mbox{cm}^{-3}$ for the highest resolution. To be more specific we summarized results of our 
simulations with different resolutions in Table \ref{table_resolution}. In addition to the magnetic field and pair density we demonstrate that the electric field may exceed vacuum value. In this case, it is a vortex electric field generated by the changing magnetic field. The power required for 
obtaining 
the same level of fields even in the case of optimal e-dipole focusing scales to 132 PW for the electric field and more than 1.2 EW for the magnetic field. We 
also 
emphasize that at a quite moderate power of 27 PW the magnetic and electric fields may rise up to 1/20 and 1/40 of the Schwinger field, respectively.
 It can also be seen that increasing resolution speeds up the process of self-compression, so the ultimate state followed by bending instability development 
is achieved at an earlier stage. This 
observation has led us to the conclusion that self-compression of the plasma column is limited by the simulation resolution and bending 
instability starts to develop at the moment of maximum compression. In fact, this means  that the limit of plasma column compression cannot be 
estimated from simulations, but this series of simulations clearly indicates the pinching nature of the pair plasma dynamics with formation of extremely dense 
electron-positron plasma objects and exceptionally high magnetic and electric fields. A very intriguing question is what the limiting plasma state is, if we do 
not observe any restrictions within the semiclassical description used in simulations. Here we may assume that they could come from the quantum physics 
requiring a quantum-mechanical approach to particle motion as well as quantum statistics for plasma description, as was proposed for counter-propagating 
electron and 
positron streams in an electron-positron collider \cite{Winterberg1979,Meierovich1982PR}. This approach may require including other QED processes in existing PIC codes to correctly resolve the dynamics of such plasma.

\begin{figure}[t!]
\centering
\includegraphics[width=8.5cm]{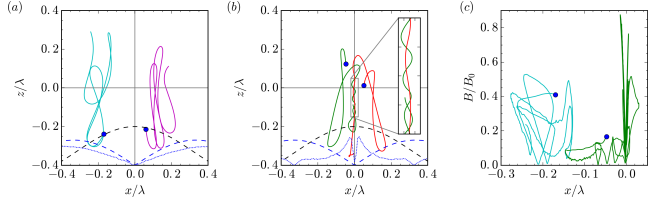}
\caption{(Color online) Typical particles trajectories for (a) 19 PW and (b) 27 PW; (c) phase plane $B$-$x$. Blue dots show 
starting points of the trajectories. Dashed lines depict electric (black) and magnetic (blue) field of a e-dipole wave. Blue dotted line shows self-consistent magnetic field at the nonlinear stage.
\label{fig4}}
\end{figure}

{\em{Particle orbits.}---}It is obvious that the azimuthal magnetic field of high amplitude generated during plasma column pinching should strongly 
alter single particle trajectories compared to the case of anomalous radiative trapping (ART)\cite{Gonoskov_ART}, mainly determining particle motion at $P<20$ 
PW. 

Indeed, the presented typical trajectories for two laser powers of 19 PW and 27 PW in Fig. \ref{fig4}(a,b) show this great 
difference. For the case of 19 PW ART trajectories corresponding to single particle motion in the field structure are close to the vacuum standing wave, when plasma do not affect field structure, and all trajectories belong to the half-plane ($x>0$ or $x<0$) and does not cross the $z$-axis. The 
picture is qualitatively different for 27 PW. Now particles are able to oscillate around the $z$-axis due to the large azimuthal magnetic field generated during 
plasma pinching. To 
emphasize this fact, we plot in Fig. \ref{fig4}(c) phase plane $B$-$x$, which clearly demonstrates the difference in the motion of these two modes. 

Since for $P=19$ PW the particles oscillate between the antinodes of electric and magnetic fields, they experience a half-maximum magnetic field and exhibit 
complex motion in the $B-x$ phase plane due to stochasticity of photon emission. In the case $P=27$ PW, the particles first drift to $x=0$ maintaining the ART pattern 
of motion in a rising but not too high magnetic field. However, when they reach the axis, the emerging strong magnetic field ($B \approx B_0$) qualitatively 
alters their trajectories\cite{Gratreau1978} and in the $B-x$ phase plane the particles show distinctive tracery looking like a set of  lengthening loops approaching 
the axis $x=0$. 
\begin{figure}[t!]
\centering
\includegraphics[width=8.5cm]{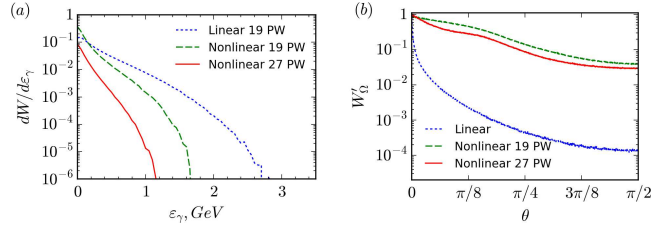}
\caption{(Color online) $\gamma$-photons: (a) spectra, (b) angular distributions for 19 PW and 27 PW e-dipole wave.
\label{fig5}}
\end{figure}

{\em{$\gamma$-ray emission.---}} As the particles' orbits, energies and momenta change drastically due to plasma pinching, another important indication of 
regime modification is alteration of the properties of the emitted photons, such as energy spectra and radiation patterns. First of all, the photon maximum 
energy changes essentially, which is clearly seen in Fig.\ref{fig5}(a), where the 
photon spectra averaged over a laser period shrink to 1 GeV, which is consistent with the new type of particle orbits, as the charged particles 
are now 
accelerated for a shorter period of time and change the direction of motion more frequently. Second, these features can also be seen in Fig. \ref{fig5}(b), where the 
radiation pattern is shown in different cases for 
comparison. In the linear regime, when the particles move predominantly along the $z$ direction, the radiation pattern is extremely narrow with the width of 1 
mrad due to the axisymmetric type of radiation, and in the non-linear regimes the angular pattern becomes more uniform with angular spread up to 0.1 rad 
relative to the $z$-axis.

In conclusion, we have shown that there are two distinctly different regimes of electron-positron plasma generation by lasers of extreme intensities. The first 
regime studied recently in Refs. \cite{Efimenko2018} takes place at a power exceeding the threshold of vacuum breakdown via QED cascading but less than about 20 
PW. At powers exceeding 20 PW, a new regime of laser-pair plasma interaction is realized, where pinching of the plasma column gives rise to 
unprecedentedly 
high pair densities. We show that this effect plays a key role in the laser-pair plasma interactions strongly affecting microscopic characteristics such as 
particle orbits and $\gamma$-ray emission, as well as macroscopic laser-plasma parameters such as pair densities and field magnitudes. It should be emphasized 
that this interaction regime leading to the generation of very small-scaled plasma-field structures is completely different from the conventional mode 
studied recently \cite{Nerush2011, Grismayer_POP2016, Jirka2016}. In this mode the pinching effect leads to the generation of magnetic and electric fields 
exceeding the incoming laser fields 
and thus may provide a novel pathway
for approaching the Schwinger field or even for overcoming this limit. Another intriguing consequence of our modeling is that the ultrahigh 
density quantum pair plasma can be produced by lasers through vacuum breakdown.  

This work was supported by the Russian Science Foundation (Grant No. 16-12-10486). 

%
\end{document}